\documentclass[a4paper]{article}
\pdfoutput=1
\usepackage{18lomcon}        % Proceedings volume layout metrics
\usepackage{cite}             % Smart range citations
\usepackage{epsfig}           % Encapsulated PostScript figure inclusion
\usepackage{epstopdf}

\newcommand{\bgfree}      {\lq back\-ground-free\rq }
\newcommand{\ctsleg}      {cts/(FWHM$\cdot$t$\cdot$yr)}

\newcommand{\zctsper}     {{10$^{-2}$~cts/(keV$\cdot$kg$\cdot$yr)}}
\newcommand{\dctsper}     {{10$^{-3}$~cts/(keV$\cdot$kg$\cdot$yr)}}

\newcommand{\tyr}         {{t$\cdot$yr}}
\newcommand{\kgyr}        {{kg$\cdot$yr}}

\newcommand{\epos}         {{$\cal{E}$}}

\newcommand{\cum}         {{m$^3$}}

\newcommand{\gesix}       {{$^{76}$Ge}}
\newcommand{\kfz}       {{$^{40}$K}}
\newcommand{\kft}       {{$^{42}$K}}
\newcommand{\arfz}       {{$^{40}$Ar}}

\newcommand{\qbb}         {{Q$_{\beta\beta}$}}
\newcommand{\tzn}         {{T$^{0\nu}_{1/2}$}}
\newcommand{\gerda}       {\textsc{Gerda}}
\newcommand{\majo}        {\textsc{Majorana}}
\newcommand{\majod}       {\textsc{Majorana Demonstrator}}
\newcommand{\igex}        {\textsc{Igex}}
\newcommand{\hdm}         {\textsc{HdM}}

\newcommand{\leg}         {{\mbox{\textsc{LEGEND}}}}

\newcommand{\tnbbd}        {{$2\nu\beta\beta$\,decay}}
\newcommand{\znbb}        {{$0\nu\beta\beta$}}
\newcommand{\znbbd}       {{$0\nu\beta\beta$\,decay}}
\newcommand{\lngs}        {{\textsc{Lngs}}}
\newcommand{\etal}        {\textit{et\,al.}}

\newcommand{\PR}         {Phys.~Rev.}
\newcommand{\PRL}         {Phys.~Rev.~Lett.}
\newcommand{\PL}         {Phys.~Lett.}

\begin{document}

%%%%%%%%%%%%%%%%%%%%%%%%%%%%%%%%%%%%%%%%%%%%%%%%%%%%%%%%%%%%%%%%%%
% The preamble of the paper
%%%%%%%%%%%%%%%%%%%%%%%%%%%%%%%%%%%%%%%%%%%%%%%%%%%%%%%%%%%%%%%%%%

%\title{Status and prospects of the search for neutrinoless double beta decay of $^{76}$Ge}
\title{STATUS AND PROSPECTS OF THE SEARCH FOR NEUTRINOLESS DOUBLE BETA DECAY OF $^{76}$Ge}

\author{Karl Tasso Kn\"opfle  for the \textsc{Gerda} collaboration \email{ktkno@mpi-hd.mpg.de}
}

\affiliation{Max-Planck-Institut f\"ur Kernphysik, Heidelberg, Germany}

\date{Dec 1, 2017}
\maketitle

%%%%%%%%%%%%%%%%%%%%%%%%%%%%%%%%%%%%%%%%%%%%%%%%%%%%%%%%%%%%%%%%%%
% The preamble of the paper
%%%%%%%%%%%%%%%%%%%%%%%%%%%%%%%%%%%%%%%%%%%%%%%%%%%%%%%%%%%%%%%%%%

\begin{abstract}
   %Fifty years of search for neutrinoless double beta decay of \gesix :
   %this review presents   
   This paper presents a review of the search for neutrinoless double beta decay of \gesix \ with emphasis on 
   the recent results of the \gerda \ experiment. It includes an appraisal of fifty years of research on
   this topic as well as  an outlook.%  to the next generation of experiments.
\end{abstract}

\section{Introduction}
Fifty years ago Fiorini and collaborators published the first paper on
the study of neutrinoless double beta (\znbb ) decay of the \gesix \ isotope,  
$^{76}$Ge$\rightarrow$$^{76}$Se\,+\,2e$^-$ \cite{fio67}.
Physics motivation was the search for lepton number violation (LNV), technical
innovation the use of a Ge(Li) crystal acting both as source and high-resolution
detector. With a lead-shielded Ge(Li) coaxial detector of 90\,g of natural germanium 
- containing 7.8\% of \gesix \ -
Fiorini et al. observed  $\sim10^{-2}$\,counts\,/(keV$\cdot$hr) around the transition energy \qbb \,=\,2039\,keV, 
but no \znbb \ signal, i.e. a sharp peak at \qbb , and
inferred after 712 hours of data accumulation a lower half-life limit of 
\tzn$>3\cdot 10^{20}$\,yr (68\%\,CL). 
%%%%%%%%%%%%%%%%%%%%%%%%%%%%%%%%%%%%%%%%%%%%%%%%%%%%%%%%%%%%%%%%%%%%%%%%%%%%%%%%%%%%%%%%%%%%%%%%%%%%%%%%%%%%
\begin{figure}[!h]
\centering
\includegraphics[height=5.2cm]{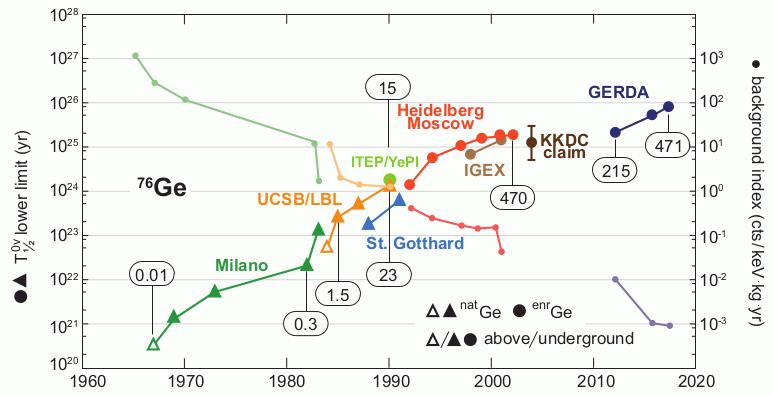}
\caption{History of \tzn \ lower limits 
(68\%\,CL, 90\%\,CL since 1991) 
and BIs from \gesix \ \znbb \ studies, 
see refs. 
\cite{fio67,bel84,cal90,reu92,vas90,kla01,aal02,kla04,ago17} and refs. therein. 
Framed numbers show exposures in mol(\gesix )$\cdot$yr.
}
\label{fig:s1}
\end{figure}
%%%%%%%%%%%%%%%%%%%%%%%%%%%%%%%%%%%%%%%%%%%%%%%%%%%%%%%%%%%%%%%%%%%%%%%%%%%%%%%%%%%%%%%%%%%%%%%%%%%%%%%%%%%%
The improvement of the \tzn \ limit in the following decades by 5 orders of magnitude (Fig.~\ref{fig:s1}) is due 
to a reduction of the background index (BI) - commonly quoted as count rate at \qbb \ normalized to energy bin and detector
mass\,M -
%  per (kg$\cdot$kg$\cdot$yr) at \qbb \ -  
and an increase of  source strength, or exposure $\cal{E}\,=\,$M$\cdot$T, with T being the run time.
%the product of mass M and run time T.
The former became possible by better shielding, including the reduction of cosmic rays 
by going underground, the latter by progress in the fabrication of larger Ge crystals. 
Both trends experienced a big boost with the introduction of Ge detectors made 
from Ge material enriched up to 90\% in \gesix \cite{vas90}.
 
This article reports mainly on the \gerda \ experiment which is expected to break the \tzn \,$=\,10^{26}$\,yr frontier 
in 2018, and on current preparations to reach \tzn \,$\ge\,10^{27}$\,yr sensitivity.
These efforts reflect the intensified interest in \znbb \ decay caused by the discovery that neutrinos have 
mass, and the theorem that the observation of \znbbd \ would establish neutrinos to have a Majorana
neutrino mass component, as predicted by various extensions of the Standard Model of particle physics (SM). 
Among the many models for LNV that of light Majorana neutrino 
exchange relates \tzn \ to an effective Majorana neutrino mass providing  access to the absolute neutrino mass scale 
and its hierarchy \cite{theo}.

\section{Present experiments}
The \gerda \ and \majo \ collaborations are currently operating a total of 65.3\,kg of 
enriched high purity Ge detectors underground. The main difference of the experiments is the
shielding against external radiation, a common feature that they discriminate \znbbd s from background 
events by their different topology: while the two electrons of a \znbbd \ would release their energy within a small detector 
volume of a few mm$^3$, background events most likely deposit energy in several locations of the detector, 
on its surface, or in adjacent detectors. Background events can thus be identified by detector-detector anti-coincidences
and, in particular, by the analysis of the time profile of the detector signal. For the latter task, both collaborations have
independently developed small read-out electrode high purity Ge detectors, p-type Broad Energy (BEGe \cite{psdg}) or point contact (PPC)
detectors \cite{psdm}, 
which exhibit not only better pulse shape discrimination (PSD) than the traditional coaxial detectors but also 
superior energy resolution, $<3\,$keV full width at half maximum (FWHM) at \qbb . To prevent any bias \gerda \ has 
adopted, first in the field, the concept of \lq blind analysis\rq : events within \qbb $\pm\,25$\,keV are cached until all analysis procedures and
cuts are finalized \cite{ago13}.
%%%%%%%%%%%%%%%%%%%%%%%%%%%%%%%%%%%%%%%%%%%%%%%%%%%%%%%%%%%%%%%%%%%%%%%%%%%%%%%%%%%%%%%%%%%%%%%%%%%%%%%%%%%%
\begin{figure}[!h]
\centering
\includegraphics[height=6.4cm]{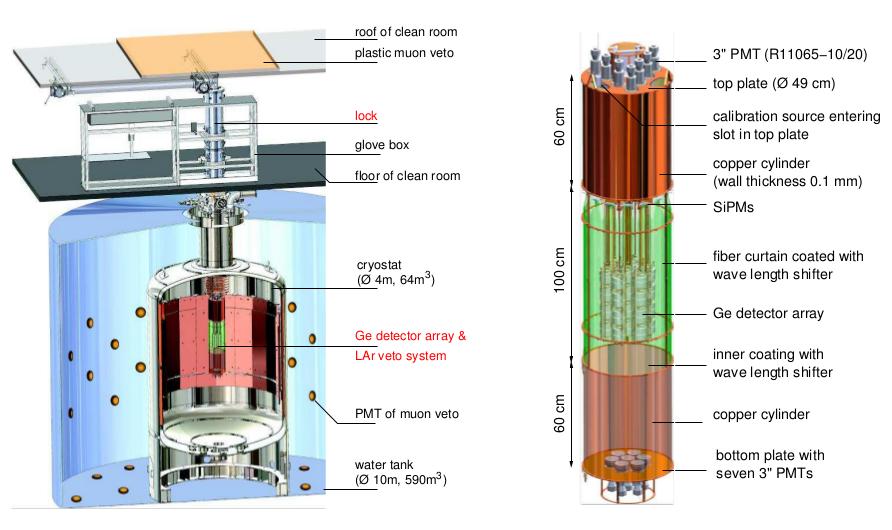}
%\includegraphics[height=6.4cm]{ktk2.eps}
%\hskip1truecm
%\includegraphics[height=6.4cm]{ktk3.eps}
\caption{The \gerda \ experimental setup (left) and the Phase\,II LAr veto system (right) \cite{gph2}. }
\label{fig:s2}
\end{figure}
%%%%%%%%%%%%%%%%%%%%%%%%%%%%%%%%%%%%%%%%%%%%%%%%%%%%%%%%%%%%%%%%%%%%%%%%%%%%%%%%%%%%%%%%%%%%%%%%%%%%%%%%%%%%

\subsection{The \gerda \ experiment}
The Germanium Detector Array (\gerda ) experiment \cite{gph1} is located at the INFN Laboratori Nazionali del Gran
Sasso (\lngs ) below a rock overburden of $\sim$3500\,m water equivalent.
The innovative feature of \gerda \ is the operation of bare Ge detectors in an ultra-pure cryogenic liquid, 
liquid argon (LAr), that serves both as cooling and shielding medium. The LAr cryostat is enclosed by a
tank filled with ultra-pure water (Fig.~\ref{fig:s2}) as additional shield against external radiation 
and as medium for a Cherenkov veto system against muons. % instrumented with 66 photomultiplier tubes (PMTs).
A clean room on top of cryostat and water tank houses a glove box and lock for assembly and deployment of 
the Ge detectors.    

In Phase I of \gerda \ (Nov 11 - May 13) 
the detector array consisted of 15.6\,kg of refurbished semi-coaxial Ge detectors from the former
\hdm \ \cite{kla01} and \igex \ \cite{aal02} collaborations and of 3.0\,kg of BEGe detectors, all enriched to 86\% in \gesix .
The background goal of BI\,=\,\zctsper \ was reached. 
With $\cal{E}$\,=\,21.6\,\kgyr \ no \znbb \ signal was observed and a new 90\% CL limit of \tzn\,$>\,2.1\cdot10^{25}$\,yr 
was set (median sensitivity $2.4\cdot10^{25}$yr) that excluded the claim of observation by part of the
\hdm \ collaboration \cite{kla04} with 99\% probability \cite{ago13}.

Phase\,II of \gerda \ started in Dec 2015 with the aim to improve the half-life sensitivity beyond 10$^{26}$\,yr. 
To achieve this at the exposure of $\sim$100\,\kgyr \ within reasonable time, e.g. 3 years of running, the detector 
mass has been doubled by augmenting the enriched BEGe detector mass to 20\,kg (30 pcs). Simultaneously, 
the BI had to be reduced by a factor  10 to \dctsper \ to stay within the \lq background-free\rq \ regime, i.e. 
a mean expected background of $<$1 in the region of interest (ROI), \qbb \,$\pm0.5\cdot$FWHM. 
This warrants the sensitivity to scale about linearly with exposure $\cal{E}$ while at larger backgrounds it scales
as $\sqrt{\cal{E}}$ (Fig~\ref{fig:s4}, left). 
The background reduction is achieved by various improvements \cite{gph2}: by the superior PSD performance of
the BEGe detectors, by new low-mass cables and detector mounts of improved radio-purity, and by the instrumentation of the
LAr surrounding the detector array with light sensitive sensors. The latter allows the detection of LAr scintillation
light using photomultiplier tubes (PMTs) and a fiber curtain read out by SiPMs (Fig. 2, right) thus creating an effective
LAr veto system against background events.
%The background reduction is tackled in various ways \cite{gph2}: by the superior PSD performance of the BEGe detectors, 
%by new low-mass cables and detector mounts of improved radio-purity, and, significantly, by the instrumentation of the 
%LAr surrounding the detector array for the detection of LAr scintillation light with photomultiplier tubes (PMTs) and a 
%fiber curtain read out by SiPMs (Fig.~\ref{fig:s2}, right) creating thus an effective LAr veto system against background events.
  
\begin{figure}[!t]
\centering
\includegraphics[width=11.6cm]{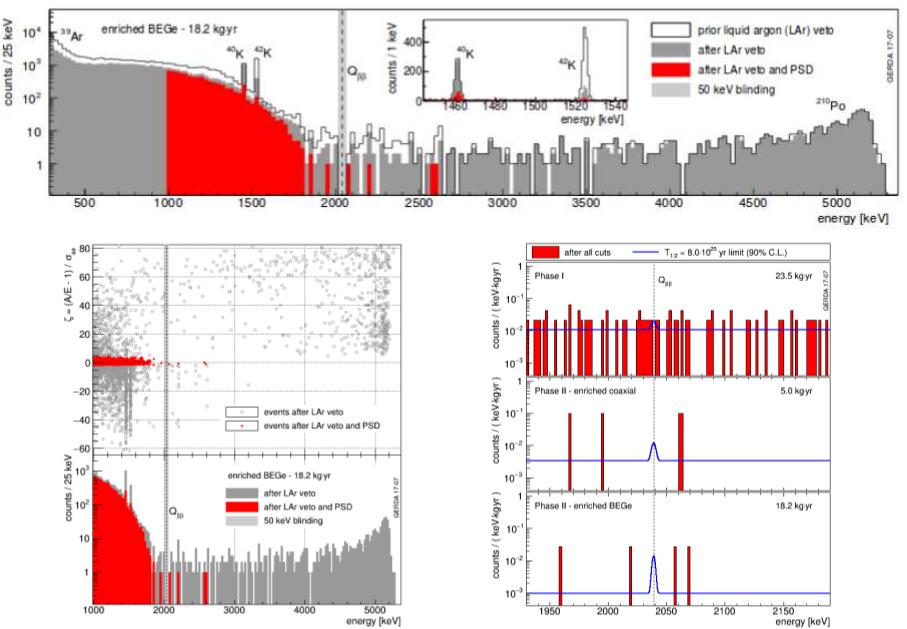}  
%\includegraphics[height=3.3cm]{ktk4.eps}   %3.2cm?
%\hskip1truecm
%\includegraphics[height=5.1cm]{ktk5.ps}
%\hskip1truecm
%\includegraphics[height=5.1cm]{ktk6.ps}
\caption{Top: Phase\,II energy spectra obtained with the BEGe detectors after indicated cuts. The energy region of
the $^{40}$K and $^{42}$K lines is shown enlarged in the inset. The vertical band indicates the blinded region
about \qbb . 
Prominent features are below 500\,keV the tail of the $^{39}$Ar\,$\beta$\,spectrum,
between 600\,-\,1600\,keV the broad structure from \tnbbd s, individual $\gamma$ lines between
400\,-\,2650\,keV, and $\alpha$\,structures above 2650\,keV, predominantly due to $^{210}$Po. -
Bottom: Scatter plot of the PSD parameter $\zeta =(A/E-1)/\sigma_{A/E}$ vs energy for all Phase\,II BEGe detectors,
and the corresponding energy spectra after indicated cuts (left). 
Energy spectra in the analysis window with 2\,keV binning obtained in Phase I and II after all cuts.
The blue lines show a fit of a constant background and a hypothetical \znbb \ signal corresponding
to the 90\% CL limit of \tzn\,$>8\cdot10^{25}$\,yr \cite{ago17} (right).}
\label{fig:s3}
\end{figure}

Fig.~\ref{fig:s3} shows the energy spectra obtained with the BEGe detectors after $\cal{E}$\,=\,18.2\,\kgyr \ and 
indicated cuts. 
%%%%%%%%%%%%%%%%%%%%%%%%%%%%%%%%%%%%%%%%%%%%%%%%%%%%%%%%%%%%%%%%%%%%%%%%%%%%%%%%%%%%%%%%%%%%%%%%%%%%%%%%%%
%%%%%%%%%%%%%%%%%%%%%%%%%%%%%%%%%%%%%%%%%%%%%%%%%%%%%%%%%%%%%%%%%%%%%%%%%%%%%%%%%%%%%%%%%%%%%%%%%%%%%%%%%%%%
In the analysis window between 1930 and 2190\,keV, the spectrum is composed of degraded $\alpha$ particles, 
Compton scattered $\gamma$'s from $^{208}$Tl and $^{214}$Bi decays, and $^{42}$K $\beta$ decays at the detector
surface.
%The performance of the LAr veto is clearly illustrated by the \kfz \ and \kft \ lines
The power of the LAr veto is illustrated by the \kfz \ and \kft \ lines
at 1461\,keV and 1525\,keV. 
Following electron capture, the line of the 1461\,keV transition in \arfz \ is unaffected by the LAr veto since 
no energy is deposited in the LAr. On the other hand, the 1525\,keV transition follows a $\beta$ decay which deposits 
up to 2\,MeV in the LAr;  thus the corresponding line can be suppressed by $\sim$\,80\%.

The PSD for BEGe detectors is based on a single parameter determined by the 
ratio A/E where A is the maximum of the detector current pulse and E the total energy \cite{psdg}. Fig.~\ref{fig:s3} 
also shows a scatter plot of the PSD parameter $\zeta\,=(A/E-1)/\sigma_{A/E}$ and its projection on the energy scale.
Accepted $\beta\beta$ candidates at $\xi=0$, mostly \tnbbd s, are shown in red; their survival fraction is 
$(85\pm2)$\%.
The two K peaks and Compton scattered events located at $\xi<0$ are easily cut, like the $\alpha$ events at $\xi>0$. 

After all cuts a BI of $1.0^{+0.6}_{-0.4}\cdot$\dctsper \ is deduced confirming a former
result from lower exposure \cite{nat}. \gerda \ is thus the first experiment in the field that will stay background-free up to its
design exposure.  
Assuming in the analysis window a flat background and a Gaussian signal at \qbb \ (and excluding 10\,keV wide intervals around
two known $\gamma$ lines), a combined unbinned profile 
likelihood fit to the Phase I and II spectra after all cuts (Fig.~\ref{fig:s3}, bottom right) yields a half-life 
limit 
for \znbbd \ of \tzn$>8\cdot10^{25}$\,yr (90\% CL, med. sensitivity $5.8\cdot10^{25}$\,yr).
For light Majorana neutrino exchange this limit converts to an upper limit of the effective Majorana neutrino mass of 
0.12\,-\,0.26\,eV using $g_A\,=\,1.27$ and nuclear matrix elements (NMEs) from 2.8\,-\,6.1 \cite{ago17}.  

\subsection{The \majod } 
The \majod \ (MJD) \cite{MAJ} is operating at the Sanford Underground Research Facility, South Dakota. 
Its goal is to prove the design for a 1 ton-scale experiment 
as to background level and modular design.
The MJD contains 35 PPC detectors (29.7\,kg) enriched up to 88\% in \gesix .  
The detectors are mounted in two vacuum cryostats within a traditional graded
passive Cu/Pb shield with the inner layer consisting  of ultraclean 
electroformed copper. 
The full detector array is running since August 2016. 

With 2.5(1)~keV FWHM resolution at \qbb, the MJD has achieved 
the best energy resolution of any \znbbd \ experiment \cite{MAJ1}, and its sub-keV
threshold allowed to perform sensitive tests of physics beyond the SM \cite{majosub}.
The lowest background runs (\epos\,=\,5.24~\kgyr ) yield a BI of 
$1.6^{+1.2}_{-1.0}\cdot$\dctsper , 
being compatible with the projected background level of 3\,cts/(ROI$\cdot$t$\cdot$yr)
based on the material assay. 
At $\cal{E}$\,=\,10~\kgyr \ no \znbb \ signal 
candidate has been observed which implies a lower limit of \tzn \,$>1.9\cdot10^{25}$~yr (90\% CL) \cite{MAJ1}.   
%At \epos\,=\,100\,\kgyr , the \tzn \ limit is expected to approach 10$^{26}$~yr.

\section{The next generation of experiments}
\begin{figure}[htb]
\centering
\includegraphics[height=4.9cm]{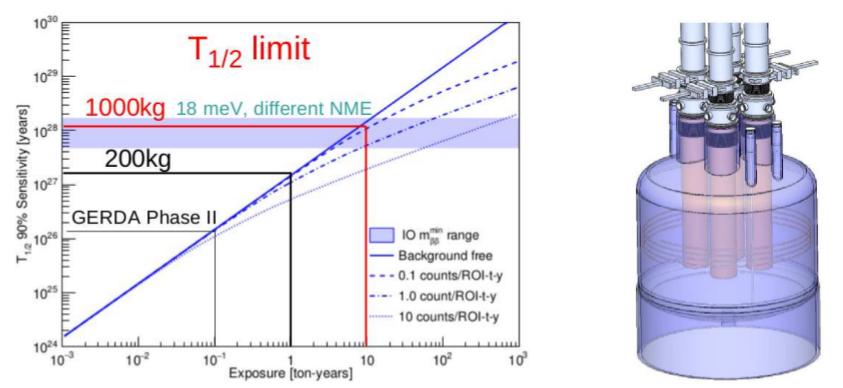}
%\includegraphics[height=4.9cm]{ktk7.ps}
%\hskip1truecm
%\includegraphics[height=4.9cm]{ktk8.ps}
\caption{Left: \tzn \ sensitivities (90\% CL) of \gerda \ Phase~II and \leg -200/1000 vs exposure; the horizontal 
band indicates the lower bound of the inverted neutrino mass ordering (IO) \cite{jdet}. 
Right: Tentative baseline design of a cryostat for \leg -1000 \cite{LEG}.}
\label{fig:s4}
\end{figure}

Already in 2005 the \gerda \ and \majo \ collaborations recognized that the ton-scale experiment
which is needed to cover the inverted mass hierarchy would call for a world-wide effort. Hence they
signed a MoU on open exchange of knowledge and technologies and declared their intentions to
merge for a ton-scale experiment combining the best features of both experiments.  
The recently formed \leg \
(Large Enriched Germanium Experiment for Neutrinoless Double Beta Decay) collaboration   
is the result of these joint efforts including new members. It proposes a staged approach \cite{LEG}:
the operation of up to 200\,kg of detectors in the existing \gerda \ infrastructure at \lngs , 
and in a next step the use of 1000\,kg of detectors in a new installation at a location 
to be still determined. Fig.~\ref{fig:s4} shows the envisaged \tzn \ limit setting 
sensitivities.    
%%%%%%%%%%%%%%%%%%%%%%%%%%%%%%%%%%%%%%%%%%%%%%%%%%%%%%%%%%%%%%%%%%%%%%%%%%%%%%%%%%%%%%%%%%%%%%%%%%%%%%%%%%%%
%%%%%%%%%%%%%%%%%%%%%%%%%%%%%%%%%%%%%%%%%%%%%%%%%%%%%%%%%%%%%%%%%%%%%%%%%%%%%%%%%%%%%%%%%%%%%%%%%%%%%%%%%%%%

For \leg -200 the \gerda \ lock and cryostat piping will be modified to
accommodate up to 200\,kg of detectors.
The major challenge will be the reduction of background to  0.6\,\ctsleg , i.e.
by a factor 5 compared to \gerda ; this is needed to stay \bgfree \ up to the
design exposure of 1\,\tyr \ where the sensitivity of 10$^{27}$\,yr is reached. Envisaged 
improvements include:
(i) the use
of PPC detectors of larger mass, 1.5\,-\,2\,kg, which becomes possible with the
novel inverted coaxial PPC detector type \cite{cooper}; one of these
devices will substitute 2-3 BEGe detectors, keeping the excellent PSD properties
while reducing the number of detector holders, cables and electronic channels by the 
same factor;
%In addition, the outer dead layer of the new Ge detectors might be increased from 
%0.75\,mm to 1.25\,mm which will reduce the background from $^{42}$K $\beta$\,decays by a 
%factor of 3. 
(ii) better LAr scintillation light collection with a denser fiber curtain
for increased $^{214}$Bi rejection; 
(iii) a new design of the electronic readout chain for better noise reduction such that PSD will
become more effective; and
(iv) the overall reduction of radio-impurities close to the detector array  
by using low-mass \majod \ style components. Preparations for \leg -200 are in progress: the
PSD power of prototype inverted coaxial detectors has been verified \cite{dom}, 
the purchase of 
Ge material enriched in \gesix \ has started, and the cryostat piping 
has been redesigned. 

The \leg -1000 goal to stay \bgfree \ for 10$^{28}$\,yr sensitivity up to the exposure of 
%$\cal{E}$\,=\,
10\,\tyr \ requires a BI of 
less than $0.1$\,\ctsleg , a factor of 30 lower than achieved in \gerda . 
Several options for reaching this goal are under study, and the experience gained with \leg -200 will
help thereby. An initial baseline design (Fig.~\ref{fig:s4}, right) shows the main cryostat volume
separated by thin copper walls from 4 smaller volumes of $\sim$3\,\cum \ each. 
Each volume carries on top a shutter and lock, like in \gerda , such that 4 payloads with up to 
250\,kg detectors can be deployed separately. The 3\,\cum \ volumes might be filled with 
depleted LAr eliminating the background due to $^{42}$Ar/$^{42}K$ $\beta$\,decays.
An important R\&D effort will be the minimization of all construction materials which do not 
scintillate and hence cannot be used to veto background events.
%decays from intrinisic radio-impurities.    

\section{Conclusion}
The \gerda \ Phase II upgrade has achieved the desired background goal
of \dctsper . \gerda \ will thus run \bgfree \ up to its design exposure of 100\,\kgyr \ reaching in 2018 a 
\tzn \ sensitivity beyond 10$^{26}$\,yr. The \majod \ is expected to exhibit a similar performance. \gesix \ experiments have 
shown the best energy resolution and the lowest background in the ROI of any isotope for \znbb \ searches \cite{comp}.
This is motivation for future experiments with 200\,kg of \gesix \ and more. The 
newly formed 
\leg  \ collaboration plans a ton-scale \gesix \ experiment for probing half-lifes \tzn \ up to 10$^{28}$\,yr. 
%for covering the inverted hierarchy. 
Preparations for the first stage, \leg -200, are in progress. 

%%%%%%%%%%%%%%%%%%%%%%%%%%%%%%%%%%%%%%%%%%%%%%%%%%%%%%%%%%%%%%%%%%
% References
%%%%%%%%%%%%%%%%%%%%%%%%%%%%%%%%%%%%%%%%%%%%%%%%%%%%%%%%%%%%%%%%%%

\end{document}